# Chip Guard ECC
# An Efficient, Low Latency Method

Tanj Bennett
Avant-Gray LLC
Senior Member, IEEE

Email: tanj@avant-gray.com

## 1. Abstract

Chip Guard is a new approach to symbol-correcting error correction codes. It can be scaled to various data burst sizes and reliability levels. A specific version for DDR5 is described. It uses the usual DDR5 configuration of 8 data chips, plus 2 chips for ECC and metadata, with 64-bit bursts per chip, to support whole-chip correction reliably and with high probity (reporting of uncorrectable faults). Various numbers of metadata bits may be supported with defined tradeoffs for reliability and probity. The method should correct all bounded faults [1] of a single chip, with less than 1 in $10^{12}$ chance of failing to correct unbounded faults in one chip, or less than 1 in $10^{12}$ chance of failure to detect an uncorrected fault which affects multiple chips.

## 2. Introduction

DDR5 memory is intended to support high-capacity memory systems which require high reliability. Manufacturing tests screen out initial flaws and either map spare resources to replace the flaw or reject the chip. However, new flaws will develop over the years that a chip is in use. DDR5 chips include a single-bit error correction mechanism [2] built into each chip which likely corrects 95% of errors over the useful life of the chip. The remaining uncorrected errors are almost entirely multi-bit faults. These are caused by flaws in structures like the word lines which are drawn at the finest resolution of any feature and interact with multiple memory cells.

A modern server may be expected to operate continuously for 5 years or more in a production environment. Each CPU socket may be supported by hundreds of DRAM chips. Modern servers use an ECC mechanism which can correct any or all errors found in a single DRAM chip. This correction of multiple bits in one DRAM, combined with the in-DRAM correction of any single bit errors on the other chips, provides a high assurance of correct operation even in large memories. The operating system or hypervisor should eventually replace faulty locations with spare resources, but the ECC is essential to allow applications to continue operation until then, and to allow the data to be safely read and copied into the new spare.

There are additional fault modes which may flip bits coming from multiple chips. These are most likely transients such as glitches in the power supply or electro-magnetic interference from nearby sources acting upon circuits and wires. These rare uncorrectables should not be allowed to pass silently.

This requires systems with large memory and a high reliability requirement to use an error correction code which can correct the entire 64-bit symbol delivered by one DDR5 chip, and which can report with high probity if there is a more extreme error which is uncorrected.

## 3. Background and Motivation

A modern server may have up to 12 DDR5 channels and each DDR5 channel includes two DDR5 sub-channels which operate independently. ECC needs to work with every store and every load on every subchannel, so there is a desire for ECC logic to be small and to use minimum power per operation because many copies of this function will be used.

It is desirable for ECC logic to have low latency including when corrections must be processed. Low latency correction allows the design to use a fixed path, with no pipeline complexity caused by a different delay when correction is invoked.

The full DDR5 DIMM design provides for 512 data bits along with 128 extra bits, using 10 DRAM chips. The 128 bits are a valuable resource that could be used for metadata. Ideally the ECC method should permit other uses for some of the 128 bits while still achieving high standards for reliability and probity of ECC.

Modern servers may have good uses for metadata associated with each cache line. Uses include (but are not limited to) marking memory locations which have been invalidated by prior uncorrected errors, cache coherency hints [3], and capability bits [4].

It is quite important to detect errors which cannot be corrected. These errors may occur due to physical interference to the power or signals of multiple chips, or to effects such as Row Hammer (disturbance) effects which may become more likely with future DRAM chips. Good probity requires these uncorrectable faults should be detected with very high probability.

DRAM is a very reliable technology, but when scaled to billions of bits per chip and hundreds of chips, errors will occur. Most of these will be single bit errors which are corrected internally by DDR5 DRAMs, but some errors come from failures which affect multiple bits. The ECC should correct all faults from any one chip and detect uncorrected faults exceeding one chip.

The usual ECC for DDR4 in servers is Reed-Solomon codes [5] which do not allow for many architectural bits while maintaining probity. Extending this to DDR5 would be a poor use of the 128-bit metadata resource. A novel alternative is described here which uses a combination of parity and a signature. This approach works better as the size of ECC resource increases. With 128 bits available, uncorrected single chip errors can be reduced by a factor of $10^{12}$ compared to not using ECC, while the chance of an uncorrected multi-chip error being undetected is also reduced by $10^{12}$, and yet 16 metabits can remain available for other architectural purposes. Other metabit allocations are possible with varying reliability: the configuration with 16 metabits is discussed here, unless otherwise noted.

## 4. Method

Each DDR5 subchannel uses 10 chips, each chip storing and returning 64 bits in any transfer (Fig. 1). 8 of these chips contain the 512 bits of user data in the transfer.

The method dedicates one chip to be a simple parity, which uses 64 bits. This is a sideways parity of the first 9 chips, so if B(c,b) is bit b of chip c, then

$$\text{Parity}(b) = \text{XOR}(B(c,b) : c \in \{0..8\}) : b \in \{0..63\}$$

The Parity chip is the 10th chip.

When the fault is in a single chip, which is the fault model to be corrected, then the Parity will show a syndrome which is the same as the bits in error, but without showing which of the chips is causing the errors.

The second requirement is then to identify which chip is at fault. The failure model ("correct up to 64 bits on any one chip out of 9 chips") gives us a lot of erasure information. There are only 10 possible erasure locations, one for each chip (including the parity chip).

Location is done by constructing a pseudo-random signature of the data in the first 8 chips and storing that in the 9th chip. That chip has 64 bits, so if 16 bits are used for features other than ECC, that leaves 48 bits for the signature. The 16 feature bits are then treated like data and included in source bits constructing the signature.

The signature function can be any pseudo-random map of source bits to the signature, but some properties improve efficiency. Every data bit has a distinct map. Construction of the final signature is separable: if any set of data bits is split into parts, then

$$\text{Sig}(A,B) = \text{Sig}(A) \text{ op } \text{Sig}(B) = \text{Sig}(B) \text{ op } \text{Sig}(A)$$

$$\text{Sig}(A,B,C) = \text{Sig}(A+C) \text{ op } \text{Sig}(B)$$

where "op" is an operation chosen to recombine separately constructed signatures.

The error detection and correction process is:

- If the burst loaded from the DIMM matches in the parity reconstruction, and matches in the signature reconstruction, then the data is passed as correct and no correction is needed.
- If the signature or parity reconstruction does not match what was read from the DRAM, then the

bit flips seen in the parity reconstruction (which possibly are none) become the correction syndrome. The syndrome is used to separately flip the matching bits in each of the 10 chips. If exactly one of the 10 reconstructions generates a signature match, that flipped chip value is taken as a correction and passed through as a success. This method depends upon the separability property of the signature.
- If recalculating the parity syndrome for each chip does not correct the signature, or if 2 or more can correct the signature, then the fault is uncorrectable.

It is worth thinking through a few special cases:

- If there is no error when the data loaded (the parity matches and the signature matches) this passes through as correct values for data and for metadata bits. There is never a false negative on a good load.
- If a single data chip has a fault, and there is no other fault, then the parity flips will correct that chip and the signature calculation using the corrected chip match. This correction is accepted if it is the only chip which corrects the signature. There is a tiny but finite chance that the flip syndrome applied to another chip will also result in a 48-bit matching signature. In this case there are 2 possible corrections, so the correction fails (detected). This is explained in more detail in section 7.
- If the signature showed no error, but the parity does not match as loaded, possibly the parity chip has the fault. Another possibility is that flips in the data or metadata randomly result in a 48-bit zero-value signature. These are distinguished by running the parity syndrome to generate signatures for data and metadata. If there is no candidate other than the parity chip then parity corrects itself, and the data and metadata are safe results. If the parity flips would generate a 48-bit zero in signature when applied to one of the other 9 chips, then the fault is uncorrectable (because it aliases to the parity chip) and will be reported (detected) as such.

That states the method in general. It is better than a single symbol correction multiple symbol detection (SSCMSD) scheme [6] because the on-die ECC in DDR5 will correct single-bit errors in each chip. Now we can look at some refinements around construction and strengthening of the signature.

## 5. A Practical Fast Signature

The signature needs to approach cryptographic grade randomness while being easy and fast to calculate.

In a specific implementation the signature chosen is an XOR of a permutation of bits (the "op" is XOR). For the data chips c ∈ {0..7} 8 bits of the signature are chosen to represent each data bit:

$$S(c, b) \in P(48, 8) : c \in \{0..7\}, b \in \{0..63\}$$

Where P(48,8) is the set of all permutations of 8 bits chosen from 48. Each $S(c, b)$ is unique.

The metabits get 19 bits of signature each:

$$S(8, b) \in P(48, 19) : b \in \{0..15\})$$

All the signatures are XOR'd together to form the completed 48-bit signature.

Due to the properties of XOR, this mapping is separable. This will be used to optimize detection and correction.

This signature is fast in logic, requiring an XOR-tree 7 levels deep to deliver each signature bit. It has the required property of being separable. The signature for any one of the 10 chips may be calculated separately, and then those are XOR-ed together to form the complete signature. Only the parity syndrome is needed, not the underlying data for the chip, when checking if the signature is corrected. This means that the correction phase only runs 9 single chip correction calculations (the parity chip does not contribute to the signature) using the same parity flip pattern for each of those 9, and then compares each of those calculations to the signature syndrome. This can be done for all chips in parallel as soon as the parity syndrome is ready, with a total complexity about the same as one extra, combined, signature.

This is not the only possible signature method. For example, CRC was considered. The separability property exists although the "op" is more complicated. However, random maps with XOR result in the simplest known implementation.

## 6. Strengthening the Signature

The potential weaknesses come from coincidences where a combination of signatures can be equal to the signatures of a different set of errors - aliasing. The weaknesses of concern are:

- single chip errors which cannot be corrected due to aliasing between chips
- uncorrected errors which are not detected because the signatures of several bits combine to cancel out to zeros

The signatures are chosen from P(48, 8) which is 377,348,994 different possibilities, so these risk for even one bit aliasing is rare. Aliasing can be made a much lesser risk by exhaustive search for collisions and choosing new patterns. For a given assignment of signatures, searches are run over all the most physically likely single chip patterns. For example, one test may check all the combinations of any two DQ (data pin) failures on any chip. Each DQ has 16 data bits that can be flipped, and there are 6 possible pairings of the 4 pins on a x4 DDR chip. DDR5 vendors design their chips so that the most common multi-bit causes will be "bounded faults" which are confined to affecting just two DQs. This pattern is therefore physically more likely than other fault patterns. It comprises 6 x ($2^{32}$) or about 25.76 billion trials, repeated over 10 chips. Running 258 billion trials takes just a couple of days on a simulation. The result will be either all patterns are non-aliased, in which case the patterns chosen are very good, or an alias is found, in which case the method will replace some of the map with new permutations and restart the checks.

It is feasible to generate patterns very likely to be good, so with few trials a pattern assignment is found where bounded faults on the DRAM chips are always free of alias, always corrected.

As for the remainder of the possible faults it is also useful to run a search for aliasing with any permutations up to 10 bits (not bounded to 2 DQs) per chip, and their inverses (up to 10 bits NOT flipped). This also is quite likely to succeed but may cause some new patterns to be chosen and testing restarted.

This exhaustive search certifies that remaining uncorrectable faults can only be physically unlikely unbounded faults and must include many bit flips.

## 7. Uncorrected Single-Chip Faults

If all patterns up to 10 bits flipped or not-flipped are shown free of aliasing, the only faults which can be aliased must have between 11 and 53 bits flipped. Since each of those will have 8 pseudo-random signature bits they cause at least 88 bit-flips within the 48 bits. This approaches the cryptographic hash ideal of a 50% chance of flipping any bit in the signature. For there to be an alias between two chips it is necessary for the same 11 or more bit-flips seen in the parity syndrome in each chip to generate the same signature syndrome. As one of the chips is always the real cause, there are 9 alternatives. This means that for any single-chip error of 11 bits or more, there is roughly a 1 in $2^{44}$ chance of failing to correct, less than 1 in 10 trillion.

Uncorrected faults of a single chip are always detected, never silent. This is because the parity chip always echoes the flipped data or special bits, so they cannot be silently undetected.

In summary, DRAM chips very rarely have errors. Of those, non-bounded-fault errors are even rarer. The signature mechanism will fail to correct just 1 in $10^{13}$ of those, which will add an infinitesimal cause of uncorrectable, but detected, memory faults.

## 8. Faults in the Special Bits

The special bits, those metadata bits not used for ECC, are awarded 19 bits of signature. This is because their correction mechanism is weaker. Ordinary data bit correction can use 3 chips (the data chip, the parity chip, and the signature chip), while special bits rely on just two chips (the parity chip and the signature chip which is also the special-data chip).

As a result, there are easily predicted flip patterns which alias the 9th and 10th chips. Each special bit has a 20-bit pattern – itself and its 19 signature bits – and then the XOR combinations of these are also aliasing patterns. Thus, there are 65,535 flip patterns on the parity chip or the signature chip which are indistinguishable from each other (2 chips equally possible sites for corrections) and thus the result is flagged as an uncorrected fault.

Although these patterns may be quickly enumerated, they are still pseudo-randomly generated and thus exceedingly unlikely to be matched by any actual physical pattern of bit flips. Exhaustive search with retry of alternate values ensures none of the uncorrrectables are bounded by any 2 DQs, nor that any have less than 11 bits flipped/not-flipped. There are about 16.7 trillion 20-bit permutations in a 48-bit signature, so the likelihood of alias or zero is reduced by a similar 1 in $10^{13}$ factor as for regular data.

## 9. Multi-Chip Faults

Multi-chip faults will not be corrected but should almost always be detected (high probity), not silent.

The goal does not include correcting multi-chip faults because the DDR5 DIMM design should make them vanishingly rare when the OS or hypervisor properly replaces known faults with spare resources, which is best-practice reliability, availability, and serviceability (RAS) in modern servers. The fault models which remain that generate multi-chip faults include row-hammer or disturbance, and electrical interference. These generate arbitrary numbers and locations of bit flips within a burst, for which no practical correction is known on a standard DDR5 DIMM and channel. Our goal is therefore to detect them, so they are not silently propagated.

The most probable model for undetected multi-chip errors is N aligned pairs of bits on different chips which cancel out on parity, and which also combine to a zero signature. It is easy to see that N = 1 cannot work (because every bit has a distinct signature) and relatively simple to do an exhaustive search for larger N. A search up to N = 5 is quite practical (causing regeneration with new patterns if a zero is found) so it can be certified that up to 5 pairs will always be detected. The faults which remain have a minimum of 12 data bits (N = 6) flipped in any pattern, with 96 signature bits affected. This achieves the 1 in 10 trillion level of rarity goal. The simple approach of exhaustive search of simple combinations, backed up by starting over for the unlikely case that a zero is found, has been tried and found to be feasible.

If data bits are not paired then a minimum of 3 chips is needed (data, parity, signature) and many more bits (because signature flips would need to be lined up with parity flips). An exhaustive search of up to 6 bits has shown that practical designs exists where no low number of bit flips could cause this kind of alias.

## 10. The design goals are met

All bounded faults are correctable. A single-chip fault more extensive than a bounded fault has less than 1 in $10^{12}$ chance of not being corrected. All uncorrected single chip faults, rare as they will be, are guaranteed to be detected.

Multi-chip faults cannot be corrected and less than 1 in $10^{12}$ chance to pass undetected.

This design requires roughly 5,000 XOR2 equivalents for Store and about 11,000 XOR2 equivalents for Load, plus latching to pace the memory interface pipeling and multiplexing for choosing original or corrected results of the Load.

## 11. Other Variations

If less than 16 special bits are required, it is possible to use more than 48 bits for the signature. If only 4 special data bits are required then signatures of 10 per data bit and 19 per special bit may be mapped into a 60-bit signature and require minimal increase in the size of the logic. This will improve the quality factor to 1 in $10^{15}$.

There may be use for this approach in other DRAM systems. A size of particular interest is 64 bits of round-trip metadata available in "inline" LP-DDR5 [7] on a CPU with a 64-byte cache line. The parity-signature approach could support a 16-bit bounded and aligned corrected symbol, with a 32-bit signature, and 16 bits remaining for architectural use. There would be around a 1 in 10 million level of failure to correct a symbol, as well as around a 1 in 10 million probity (failure to detect an unbounded error).

Another possible variation is to cover a 9-chip DIMM channel for corrections to any bounded fault in adjacent DQs. Those are 32-bit sequences aligned at 16-bit boundaries. There are 3 such boundaries per chip, so there would be 27 erasure positions to be checked. The reliability of such a bounded-fault correction would be slightly less than a full chip correction, and the probity level is likely to be 1 per million not 1 per trillion, but the saving of one chip (9 instead of 10) may be reasonable for some applications.

## 12. Acknowledgements

The author would like to thank Eric Shiu, Tao Zhang, Greg Lucas, Tom Bielecki, and Raymond Xiao, at Rivos Inc., for their guidance, insights, and support during the development and verification of the Chip Guard algorithms and implementation.

## 13. Figures

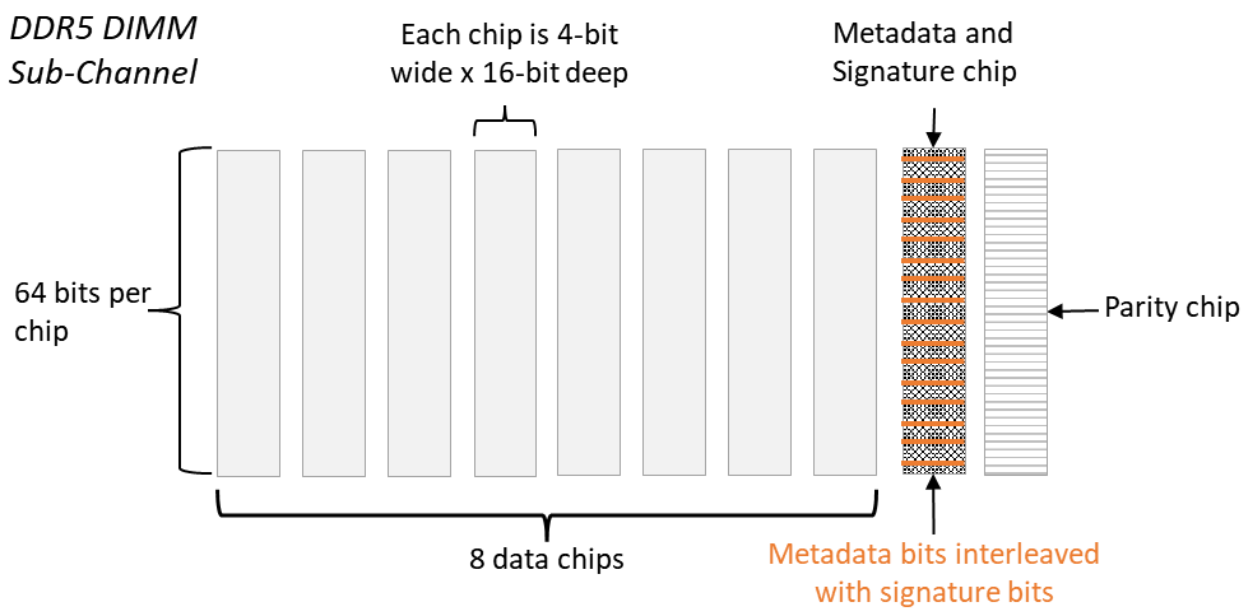

*Figure 1*

## 14. References

[1] Improving Memory Reliability by Bounding DRAM Faults: DDR5 improved reliability features
https://dl.acm.org/doi/10.1145/3422575.3422803

[2] JEDEC Publishes New DDR5 Standard https://www.jedec.org/news/pressreleases/jedec-publishes-new-ddr5-standard-advancing-next-generation-high-performance

[3] Topology and Cache Coherence in Knights Landing and Skylake Xeon Processors
https://www.ixpug.org/documents/1524216121knl_skx_topology_coherence_2018-03-23.pptx Slides 7-8

[4] Capability Hardware Enhanced RISC Instructions (CHERI) https://www.cl.cam.ac.uk/research/security/ctsrd/cheri/

[5] Wu, Y. (2015). New scalable decoder architectures for Reed–Solomon codes. IEEE Transactions on Communications, 63(8), 2741-2761. https://doi.org/10.1109/TCOMM.2015.2445759

[6] Yeleswarapu, R., & Somani, A. K. (2021). Addressing multiple bit/symbol errors in DRAM subsystem. *PeerJ Computer Science*, *7*, e359.  https://doi.org/10.7717/peerj-cs.359

[7] DRAM Error-Correcting Code Memory on the Platform,
https://docs.nvidia.com/drive/drive_os_5.1.6.1L/nvvib_docs/index.html#page/DRIVE_OS_Linux_SDK_Development_Guide/System%20Programming/sys_components_dram_ecc.html